\documentclass[elsart1p,letter]{elsart}

\usepackage{epsfig}
\def\cf4      {CF${_4}$\ }

\usepackage{lineno}

\begin{document}

\begin{frontmatter} 

\title{
Observation of the ``head-tail'' effect in nuclear recoils of low-energy neutrons 
}

\author[MIT]{D.~Dujmic\corauthref{cor}}
\corauth[cor]{Corresponding author.} 
\ead{ddujmic@mit.edu}
\author[BU]{H.~Tomita}
\author[BU]{M.~Lewandowska}
\author[BU]{S.~Ahlen}
\author[MIT]{P.~Fisher}
\author[MIT]{S.~Henderson}
\author[MIT]{A.~Kaboth}
\author[MIT]{G.~Kohse}
\author[MIT]{R.~Lanza}
\author[MIT]{J.~Monroe}
\author[BU]{A.~Roccaro}
\author[MIT]{G.~Sciolla}
\author[Brand]{N.~Skvorodnev}
\author[MIT]{R.~Vanderspek}
\author[Brand]{H.~Wellenstein}
\author[MIT]{R.~Yamamoto}

\address[BU]{Boston University, Boston, MA 02215}
\address[Brand]{Brandeis University,  Waltham, MA 02454}
\address[MIT]{Massachusetts Institute of Technology, Cambridge, MA 02139}

\begin{abstract}
Directional detection of dark matter can provide unambiguous observation of dark matter 
interactions even in the presence of background. 
This article presents an experimental method to measure the direction tag (``head-tail'') 
of the dark matter wind by detecting the scintillation light created by the elastic 
nuclear recoils in the scattering of dark matter particles with the detector material. 
The technique is demonstrated by  tagging the direction of the nuclear recoils created
in the scattering of low-energy neutrons with \cf4  in a low-pressure time-projection chamber that is 
developed by the DMTPC collaboration.
The measurement of the decreasing ionization rate along the recoil trajectory provides the direction tag
of the incoming neutrons, and proves that the ``head-tail'' effect can be observed.  
\end{abstract}


\end{frontmatter}

%
%

\section{Introduction}
\label{sec::introduction}
Searches for non-baryonic dark matter in the form of weakly interacting massive particles (WIMPs)
rely on  detection of  nuclear recoils created by the elastic scattering between a WIMP 
and the detector material. 
The current generation of experiments~\cite{DMreview}  attempt to suppress all backgrounds
to negligible levels so that any remaining events would be attributed to the WIMP signal.
However, as the size of the apparatuses increases and the sensitivity to dark matter improves, 
some irreducible backgrounds will start to appear,  
rendering a positive observation of a dark matter signal suspect. 
Examples of such backgrounds are nuclear recoils due to neutrons generated by 
cosmic rays in the rock surrounding the detector, 
or  neutrinos from the sun~\cite{Monroe:2007xp}.  
The measurement of an annual modulation of the interaction rate of dark matter 
has been suggested~\cite{amplitude modulation}, but this effect is expected to be small  (a few percent). 

An unambiguous observation of dark matter in presence of background 
is   possible by  detecting  the direction of the incoming dark matter particles.
As the Earth moves in the galactic halo with a velocity of approximately 220~km/s, 
the dark matter particles appear to  come from the 
Cygnus constellation. By measuring  the direction  of the WIMPs and correlating such measurement  with the 
position of Cygnus in the sky, an experiment can gain orders of magnitude 
in sensitivity to dark matter~\cite{directionality}. 
The determination of the vector direction of the incoming particle, 
often referred to as the ``direction tag'' or the ``head-tail'',  
is very important since it further increases the sensitivity of a directional detector
by approximately an order of magnitude~\cite{agreen}. 

The DRIFT experiment~\cite{DRIFT} pioneered the study of directional 
detection of dark matter and demonstrated the ability to reconstruct 
the direction of the incoming particles by detecting the direction of recoiling 
nuclei in a gaseous detector. However, the capability of detecting the direction tag of the incoming 
particles has not been demonstrated  by any experiment to date.

This paper demonstrates a technique to determine the direction tag (``head-tail'') 
of low-energy nuclear recoils created by dark matter particles 
by using a time-projection chamber (TPC) with optical readout developed by the DMTPC collaboration.
The projection of nuclear recoils
along the anode wires of the TPC is recorded by a charge-coupled device (CCD) camera 
imaging the scintillation photons produced during the avalanche process. 
The measurement of the direction tag relies on the fact that the 
stopping power (dE/dx) of  recoiling nuclei depends on their residual energy, 
and therefore the recoil direction can be tagged from the light distribution along the track.
The energy of the nuclear recoils created by the 
scattering of low energy neutrons or dark matter particles 
is of the order of a few keV per nucleon, well below the Bragg peak. 
Therefore a decreasing light yield along the track is expected.

%
%

\section{Experimental Setup}
\label{sec::setup}

\begin{figure}[ht]
\setlength{\unitlength}{1cm}
\begin{picture}(14,9)
\thicklines
%
%
\put(5.5,6.5){\framebox(3, 1.5)}
\put(5.5,7.25){\line(1,0){3}}
\multiput(6.5,7.1)(0.1,0){10}{\framebox(0.1,0.15)}
\put(9.5,7.1){CCD camera}
%
%
\qbezier(5,6)(7,6.25)(9,6)
\qbezier(5,6)(7,5.75)(9,6)
\put(9.5,6){Lens}
%
%
\put(6.2,5.1){Window}
\put(3,5){\framebox(8, 0.5)}
%
%
\put(0,0){\framebox(14, 5)}
%
%
\put(5,3){\cf4 (100-380~Torr)}
%
%
\put(1,4.1){Cathode grid}
\put(1,3.5){-1.5~kV}
\put(1,3.9){\line(1,0){2}}
\multiput(3,3.9)(0.075,0){108}{\circle*{0.01}}
\put(11,3.9){\line(1,0){2}}
%
%
\put(1,2.2){Grounded}
\put(1,1.8){wire plane}
\put(2.95,2.05){\vector(2,-1){0.65}}
\put(1,1.6){\line(1,0){2}}
\multiput(3.1,1.6)(0.16,0){50}{\circle*{0.03}}
\put(11,1.6){\line(1,0){2}}
%
%
\put(10.7,2.8){Anode}
\put(10.7,2.4){wire plane}
\put(10.7,2.0){+3~kV}
\put(10.5,2.05){\vector(-2,-1){1.4}}
\put(1,1.3){\line(1,0){2}}
\multiput(3.2,1.3)(0.40,0){20}{\circle*{0.06}}
\put(11,1.3){\line(1,0){2}}
%
%
\put(1,1){\line(1,0){12}}
\put(1,0.5){Grounded plate}
\put(4,0.55){\vector(2,1){0.7}}
%
%
\put(13.25,2.9){\vector(0,1){1.0}}
\put(13.25,2.4){\vector(0,-1){0.8}}
\put(12.8,2.6){\small 26 mm}
\put(13.25,0.8){\vector(0,1){0.5}}
\put(13.0,1.3){\small 3 mm}
\end{picture}
\caption{Schematics of the detector. \label{fg::experimental setup}}
\end{figure}
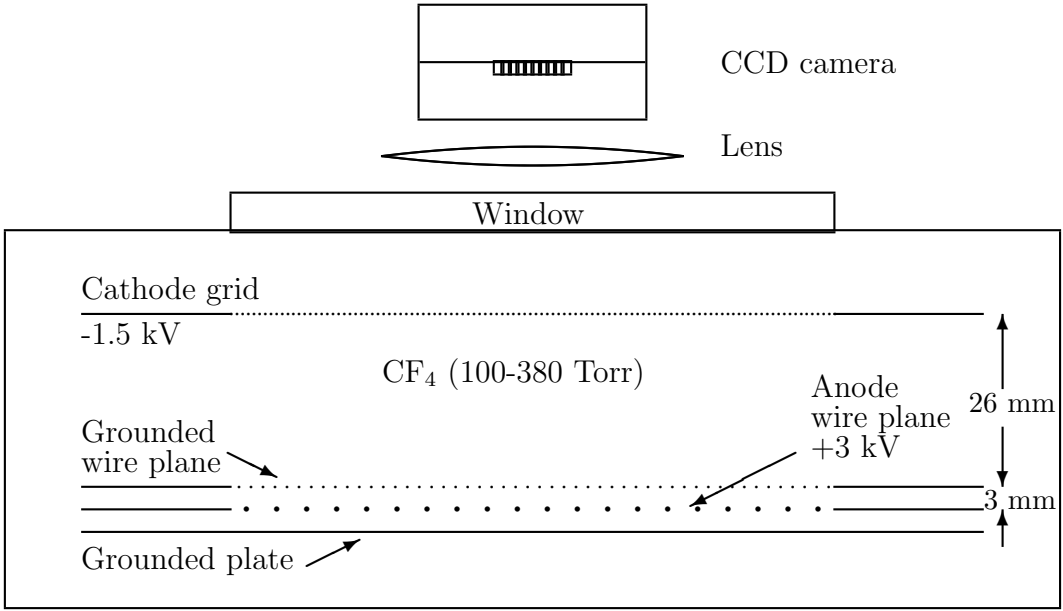

A schematic of the detector is shown in Figure~\ref{fg::experimental setup}.
The chamber utilizes   $10 \times 10~{\rm cm}^2$ wire frames. 
The  drift region between the cathode mesh and the ground 
wire plane is 2.6~cm, while the 
amplification region between the  ground  and the anode wire planes is about 3~mm. 
The pitch of the wires for the ground (anode) plane is 2~mm (5~mm) and 
the wire diameter is 50~$\mu$m (100~$\mu$m).
The chamber is filled with \cf4 at 150-380~Torr. 
The pressure is monitored with a  capacitance gauge (Inficon PCG400) in the calibration runs
with alpha particles, and a thermocouple gauge (LVG-200TC) in the nuclear scattering runs.

The scintillation light is recorded with a CCD camera manufactured by 
Finger Lake Instrumentation equipped with a $768 \times 512$ CCD chip (Kodak KAF-0401E). 
The camera has a built-in cooler that maintains the temperature in the range [-20,-18]~C to minimize  electronic noise. 
The pixel size is 9$\times$9 $\mu \rm{m}^2$. 
The photographic lens has the aperture ratio, f/\# of 1.4 and the focal length of 55~mm. 
The peak value for the quantum efficiency of the CCD chip is approximately 80\%.
The gain of the camera is measured to be $1.6$ e$^-$/ADC count.
The RMS spread of pixel yields due to ADC noise and dark current is measured to be 7 counts when all pixels are read out, 
and 25 counts when  $8 \times 8$ pixels are combined  during
the CCD readout.  Pixels that have intensity greater than 5 standard deviations from the mean dark field 
at least 10\% of the time are flagged as `hot channels' 
and excluded in the data analysis. 
ADC bias is corrected for by subtracting from each image 
the average of 100 images taken with the shutter closed. 

Neutrons used in this work are produced isotropically with 14.1~MeV energy from deuteron-triton 
reactions in a neutron generator (Thermo MF Physics A-325).  
The deuterium plasma is generated by a 3~kV voltage pulse and
accelerated through a 70~kV region toward a triton target.
The pulses are 100~$\mu$s wide and issued at a frequency of 1~kHz. 
Using the manufacturer's
specifications, the 
total isotropic flux of neutrons is estimated to be $5 \cdot 10^{7}$~neutrons/s.
The solid angle imaged with the CCD camera is $7.2 \times 10^{-4}$~sr.
%
%
%
%
%
%

\section{Calibration with $\alpha$ Source}
\label{sec::alpha}

The detector response is evaluated using 5.5~MeV $\alpha$ 
particles from a collimated $^{241}$Am source.
The chamber is filled with  \cf4  at various pressures in the range 100-380~Torr.  
The drift field is set to 580~V/cm, while 
the amplification voltage is varied between  2.1 and 4.1~kV. Images are taken sequentially with 1~second exposure time.

\begin{figure}[hb]
\center
\includegraphics[width=10cm]{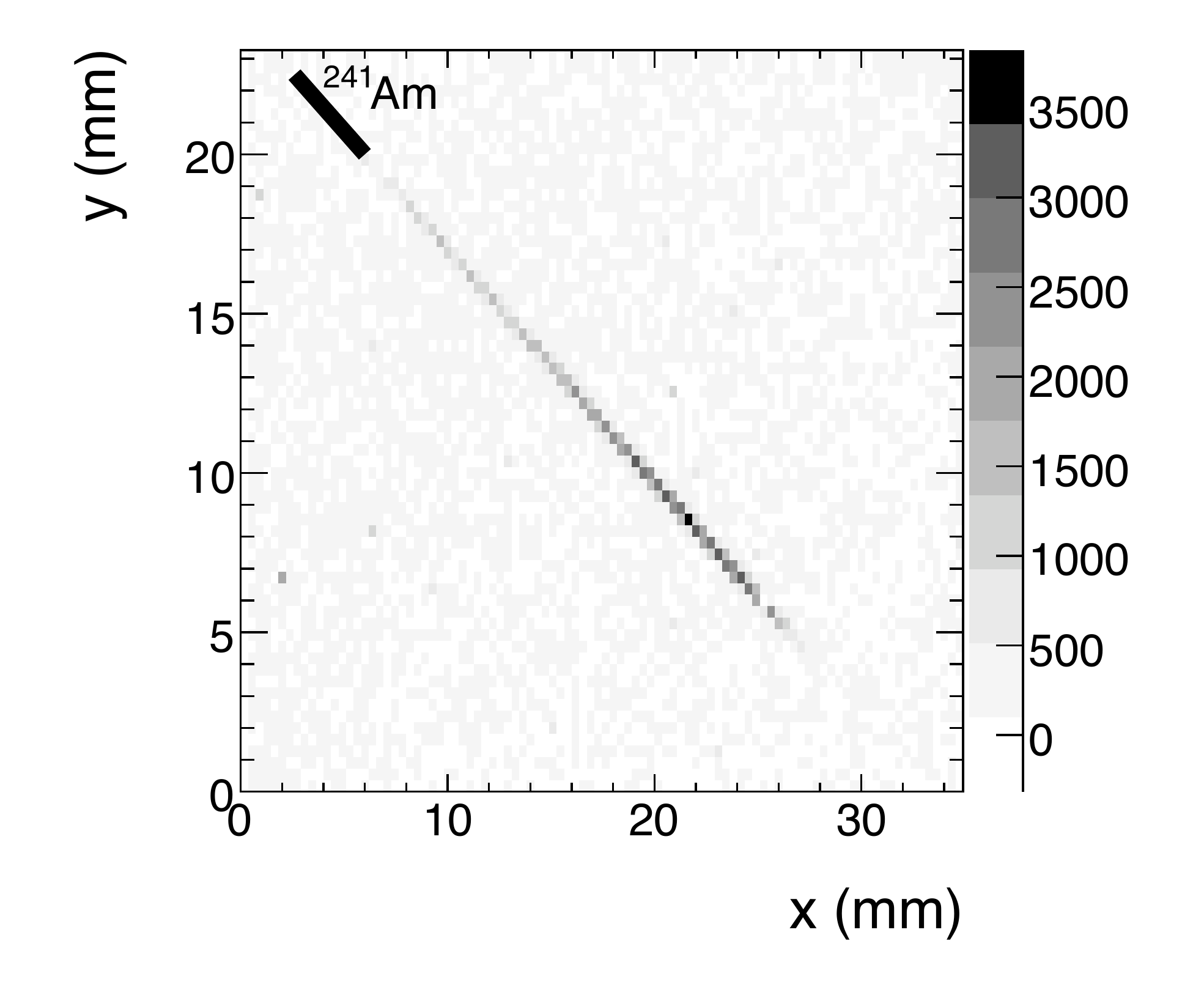} \\
\includegraphics[width=10cm]{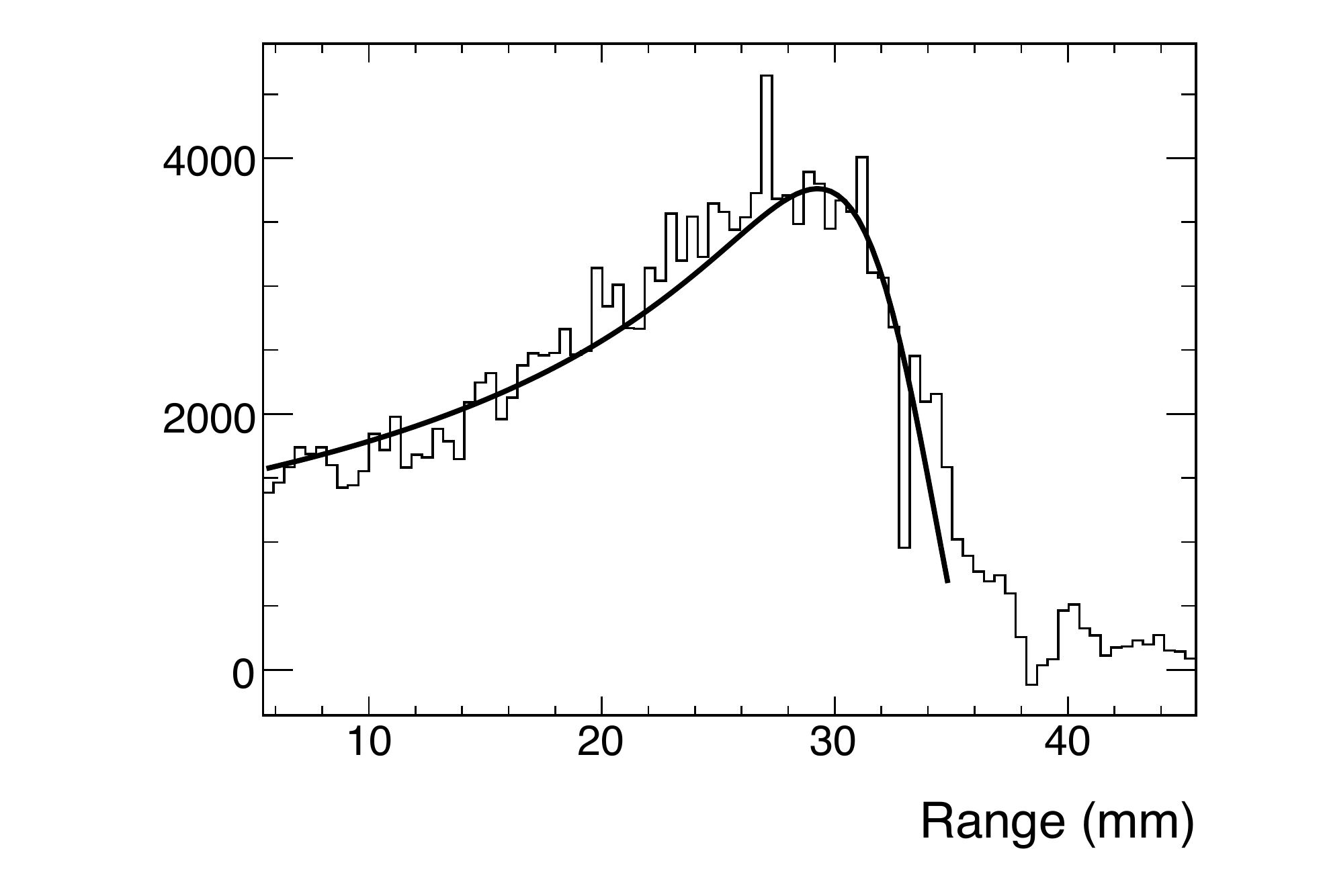} 
\caption{Top plot: accumulation of 12 CCD images of $\alpha$ tracks traveling parallel to the anode wires in \cf4 at 280~Torr. 
 The source is placed on the top-left corner. 
 Bottom plot: light yield measured in ADC counts vs. range of the recoil  in data (histogram). 
The line shows the fit to the distribution 
obtained by using the stopping power and straggling from the SRIM simulation package~\cite{srim}. 
\label{fg::alphas parallel to wires}}
\end{figure}

Figure~\ref{fg::alphas parallel to wires} shows the accumulation of 12  images of $\alpha$ tracks
from a source placed in the upper-left corner of the CCD field of view 
with its collimator pointing toward the lower-right corner.  
The anode wires are oriented parallel to the $\alpha$ source. 
As the particle travels in the medium, 
the intensity of the scintillation light increases until it reaches a maximum  
corresponding to the Bragg peak, 
and then decreases toward the end of the track. 
The longitudinal scintillation profile 
is shown in the lower plot of Figure~\ref{fg::alphas parallel to wires}. 
This profile can be described with the energy loss
due to ionization and excitation of gas molecules. 
This assumption  is verified by 
fitting the  distribution  with the stopping power curve obtained with the SRIM simulation.
The result of such a fit is shown in the same plot.

The range of $\alpha$ particles in \cf4 is obtained by  varying the end-point of the track in the fit. 
Images at pressures ranging from 280 to 380~Torr  are taken in 20~Torr increments. 
The range measured in data ($R_{data}$) is compared with values predicted by the simulation  ($R_{SRIM}$). 
The agreement between data and simulation 
is found to be  within the experimental errors, $R_{data}/R_{SRIM}=0.85 \pm 0.11(stat)\pm0.10(syst)$. 

\begin{figure}[hb]
\center
\includegraphics[width=11cm]{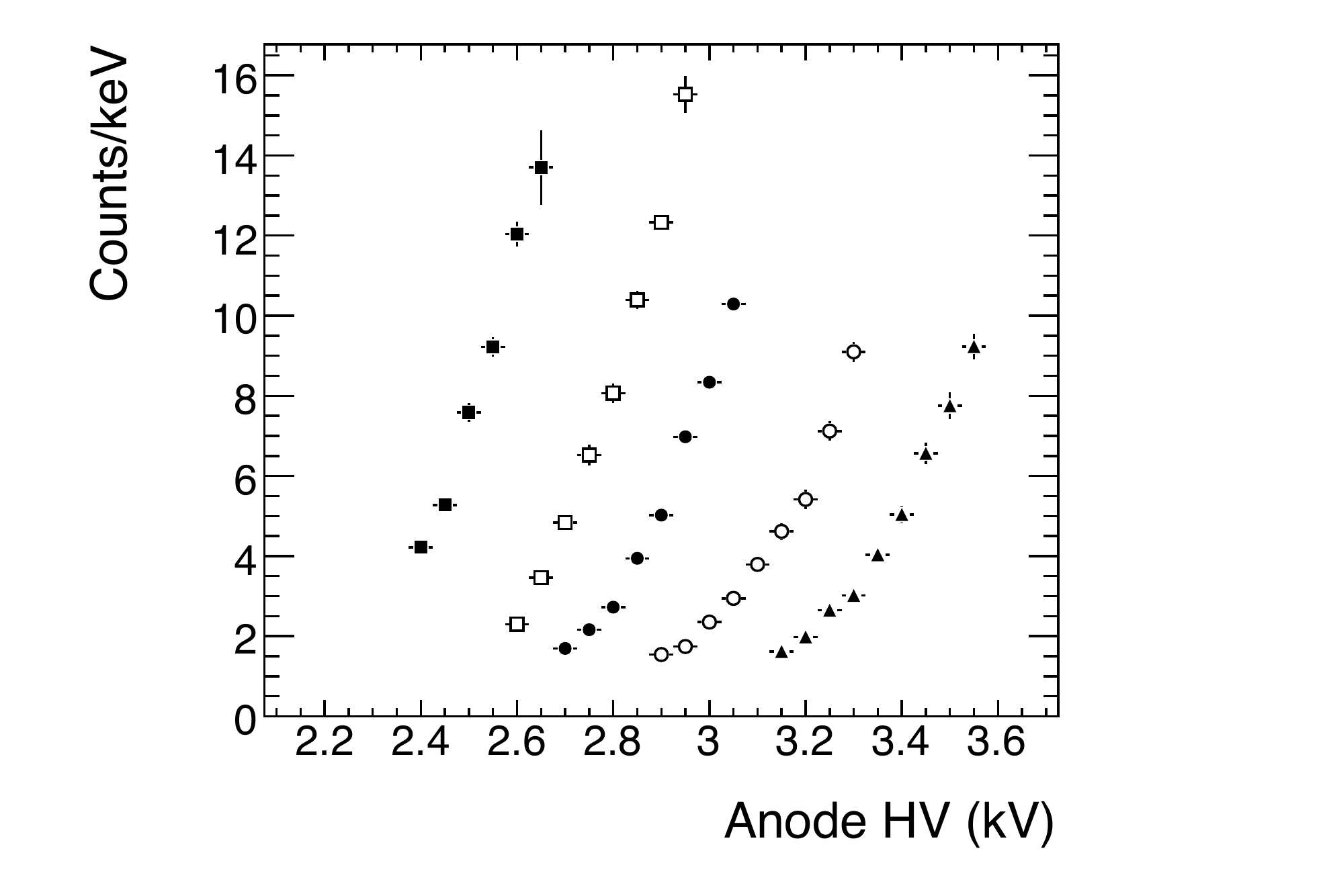} 
\caption{Light intensity normalized to the energy loss at a wire plotted against amplification
voltage. Marker style denotes the \cf4 pressure (from left to right): 150, 200, 250, 300, 350~Torr.
\label{fg::counts_vs_hv}}
\end{figure}

The gain of the detector is determined by measuring 
the intensity of the scintillation light recorded by the CCD camera 
as a function of the energy deposited in the detector 
by $\alpha$ particles oriented perpendicular to the anode wires. 
SRIM is used to estimate the total energy deposited at each wire.  
Figure~\ref{fg::counts_vs_hv} shows the average yield in the CCD normalized to the expected energy 
loss over a wire for different pressures and amplification voltages.

The signal spread due to electron diffusion through the gas is one of the critical parameters in the design of a detector for dark matter search.
Since the expected range of fluorine recoils in \cf4 gas at 50 Torr is a few millimeters,
the effect of the diffusion must be contained to well below 1~mm. 
This requirement defines the maximum drift length of the detector.

\begin{figure}[hb]
\center
\begin{tabular}{ll}
\includegraphics[width=7.8cm]{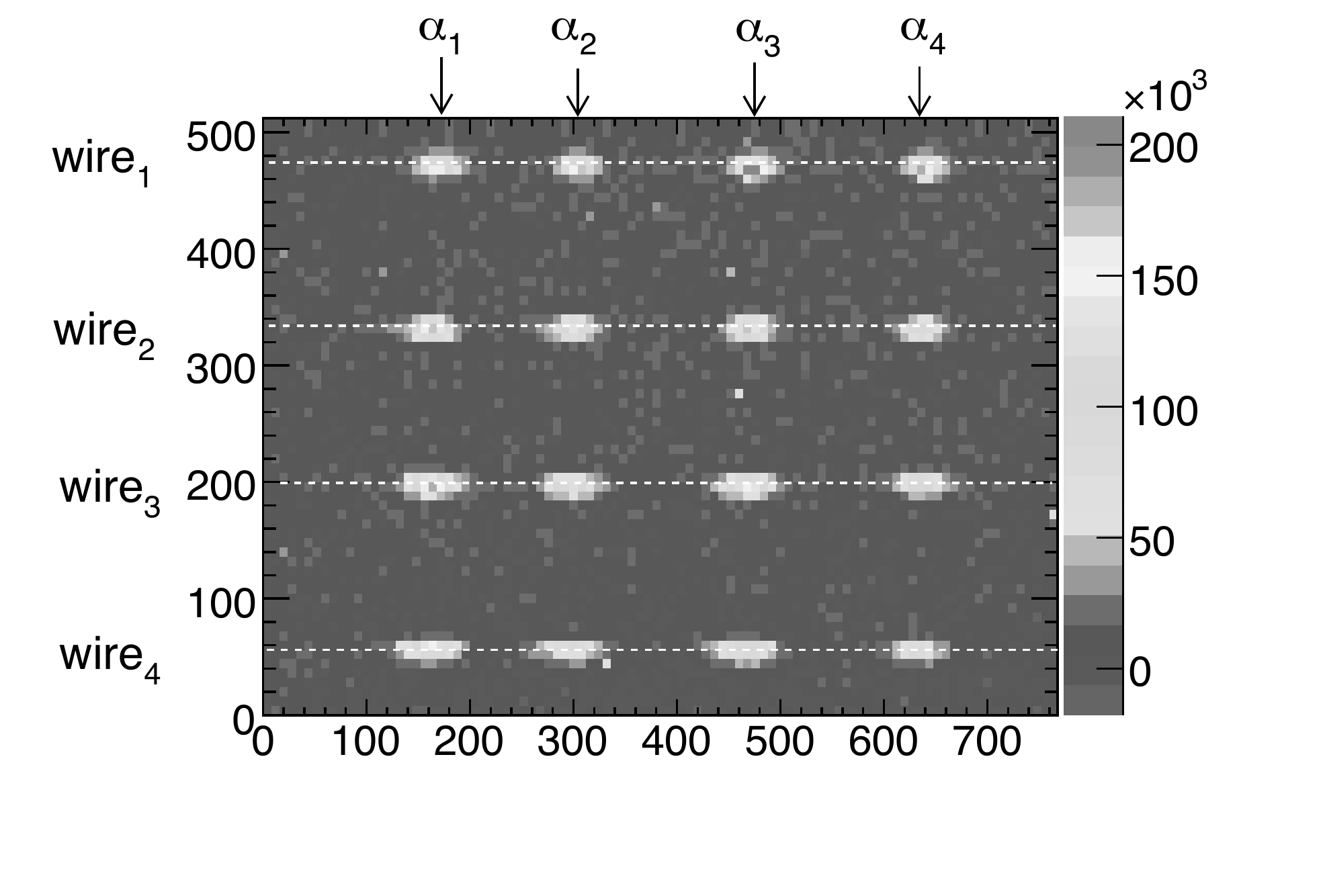} &
\includegraphics[width=7.0cm]{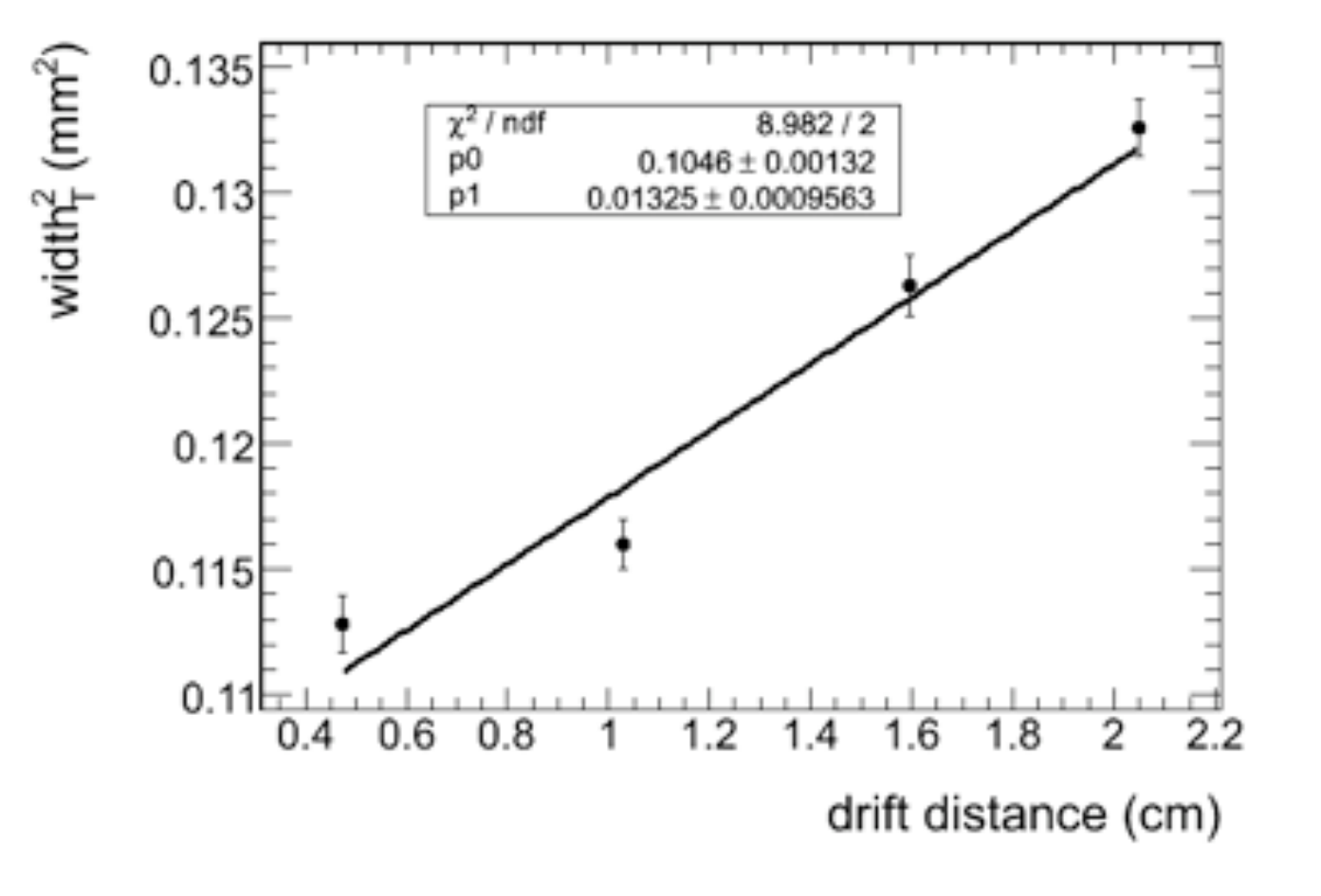}
\end{tabular}
\caption{The left plot shows the accumulation of 1700 CCD images of $\alpha$ tracks produced by 
four  sources placed at different drift distances (0.48, 1.02, 1.6, 2.05~cm). 
The $\alpha$ particles are traveling perpendicularly to the wires with a pitch of 5~mm. 
Note that accumulated hit images get wider as alpha tracks move away from a source due to imperfect collimation. 
The diffusion is measured as a change in the  width between sources for single tracks.
The right plot shows the linear dependence between the square of the signal width and the drift distance.
\label{fg::four sources}}
\end{figure}

The diffusion parameters are measured using four $\alpha$ sources placed perpendicular to the anode wires 
at different heights ($\Delta z$) above the grounded plane of the amplification region. 
 Figure~\ref{fg::four sources}  shows the accumulation of 250 images of $\alpha$ tracks coming from the four sources. 
The activity of the sources is small enough that each image contains only one $\alpha$ track.
Since there is no electric field parallel to the wires, 
the width of the signal along the wires is a measure of the electron
diffusion in the drift and amplification regions. 
Taking the average width from all wires at a pressure of approximately 220~Torr, 
the resolution as a function of the drift distance is measured to be 
$\sigma_T [\mu \rm{m}]= (324 \pm 2) \oplus (36 \pm 4) \sqrt{\Delta z [\rm{mm}]}$,
where the errors are statistical. 
These measurements show that 
the maximum drift distance in this detector should be limited to 
25~cm, for which the resolution is ($670 \pm 20$)~$\mu \rm{m}$.

%
%
%
%
%
%

\section{Neutron Beam Results}
\label{sec::results}

Nuclear recoils produced in dark matter scattering are simulated by using a 14.1~MeV neutron beam.   
Figure~\ref{fg::energy angle for neutrons, wimps} compares the energy and scattering angle of the nuclear 
recoils generated by a 200 GeV WIMP (solid line) and by 14.1~MeV neutrons (dashed line). 
Overall the distributions are similar. 
Note that the lower energy of the recoils generated by WIMPs causes shorter track lengths. This 
effect is compensated by a much better alignment of the recoil momentum with the WIMP direction.

The neutrons enter the tracking chamber through a small opening in a 19~cm thick concrete shield. 
The wires of the tracking chamber are aligned with the direction of the neutron beam.

\begin{figure}[hb]
\center
\begin{tabular}{ll}
\includegraphics[width=6.5cm]{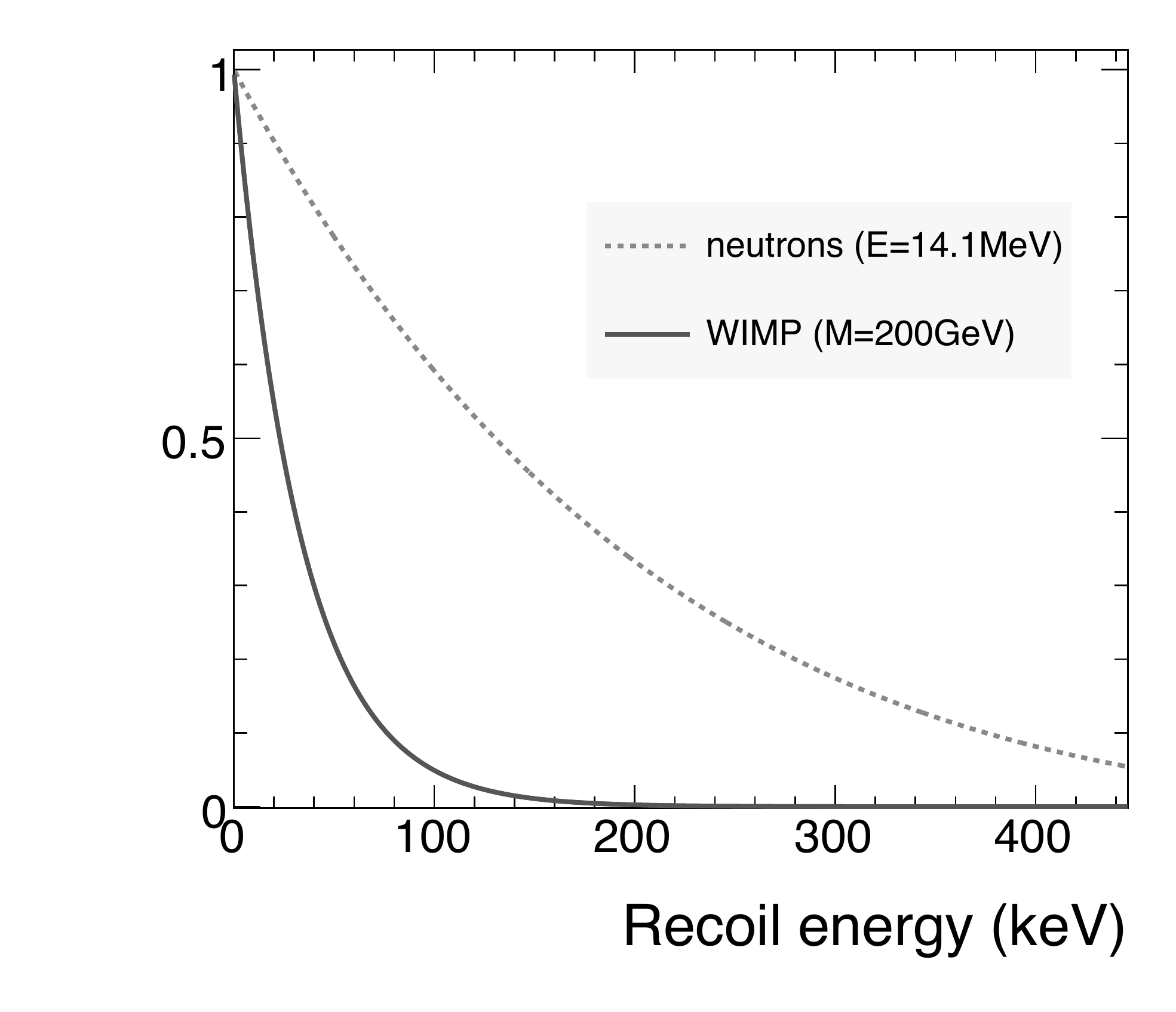} &
\includegraphics[width=6.5cm]{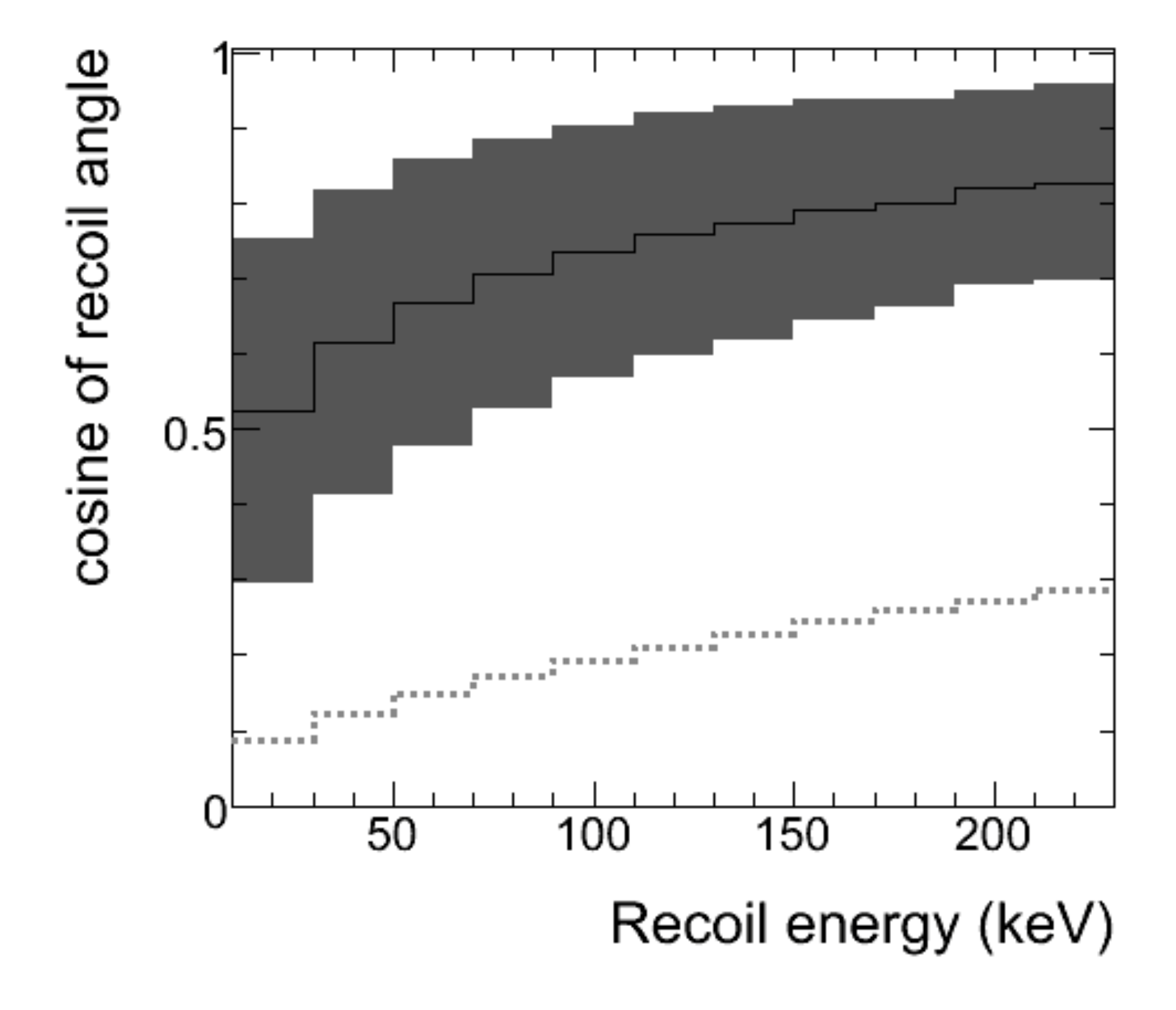} 
\end{tabular}
\caption{Energy spectrum and distribution of recoil angle vs. energy 
for nuclear recoils generated by 
the scattering of 14.1~MeV neutrons (dashed line) and 200~GeV WIMPs (solid line)
 with fluorine nuclei. 
The recoil angle is defined as the angle between the direction of the 
nuclear recoil and the direction of the incoming
particle. 
\label{fg::energy angle for neutrons, wimps}}
\end{figure}

The tracking chamber is operated with cathode voltage of -1.5~kV, 
amplification anode voltage of +3~kV, and \cf4 pressure of 200-250~Torr. 
Sequential 1-second exposures are taken with the CCD camera, imaging a region of 1~cm$^2$ area.  
Only a few percent of the exposures are expected to contain the signature of a fluorine recoil
since the total cross section for scattering of neutrons on fluorine is 1.75 b, and 
the cross section for elastic scattering is 0.9~b~\cite{endf}. 
The dominant background comes from the F(n,n+$\alpha$)N process, 
with the cross section estimated at 0.45~b.  
However, the $\alpha$ particles produced in this process 
are more energetic and have smaller energy 
loss than fluorine recoils, so they can be easily rejected.

\begin{figure}[hb]
\center
\begin{tabular}{ll}
\includegraphics[width=6.5cm]{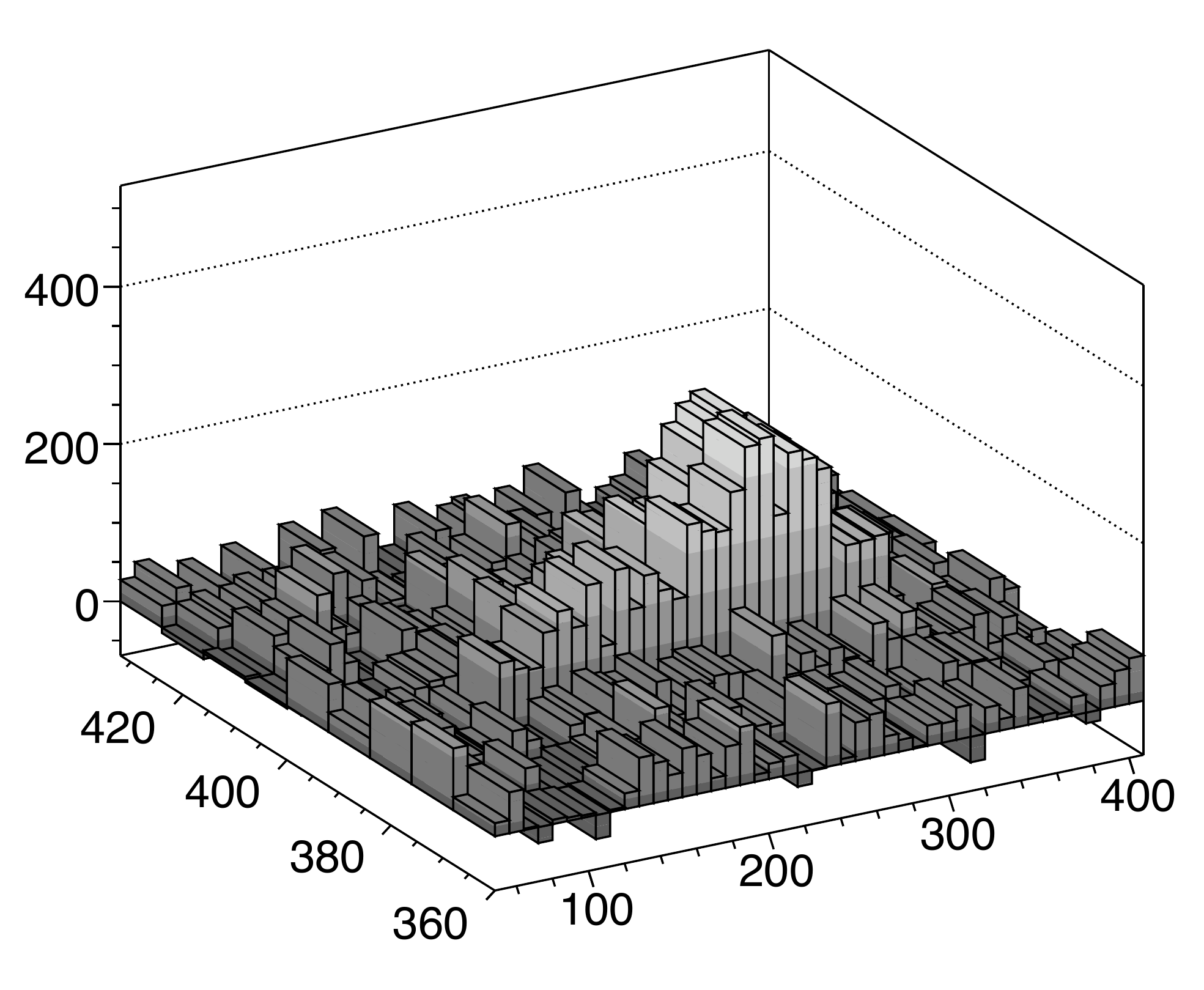} & \includegraphics[width=6.5cm]{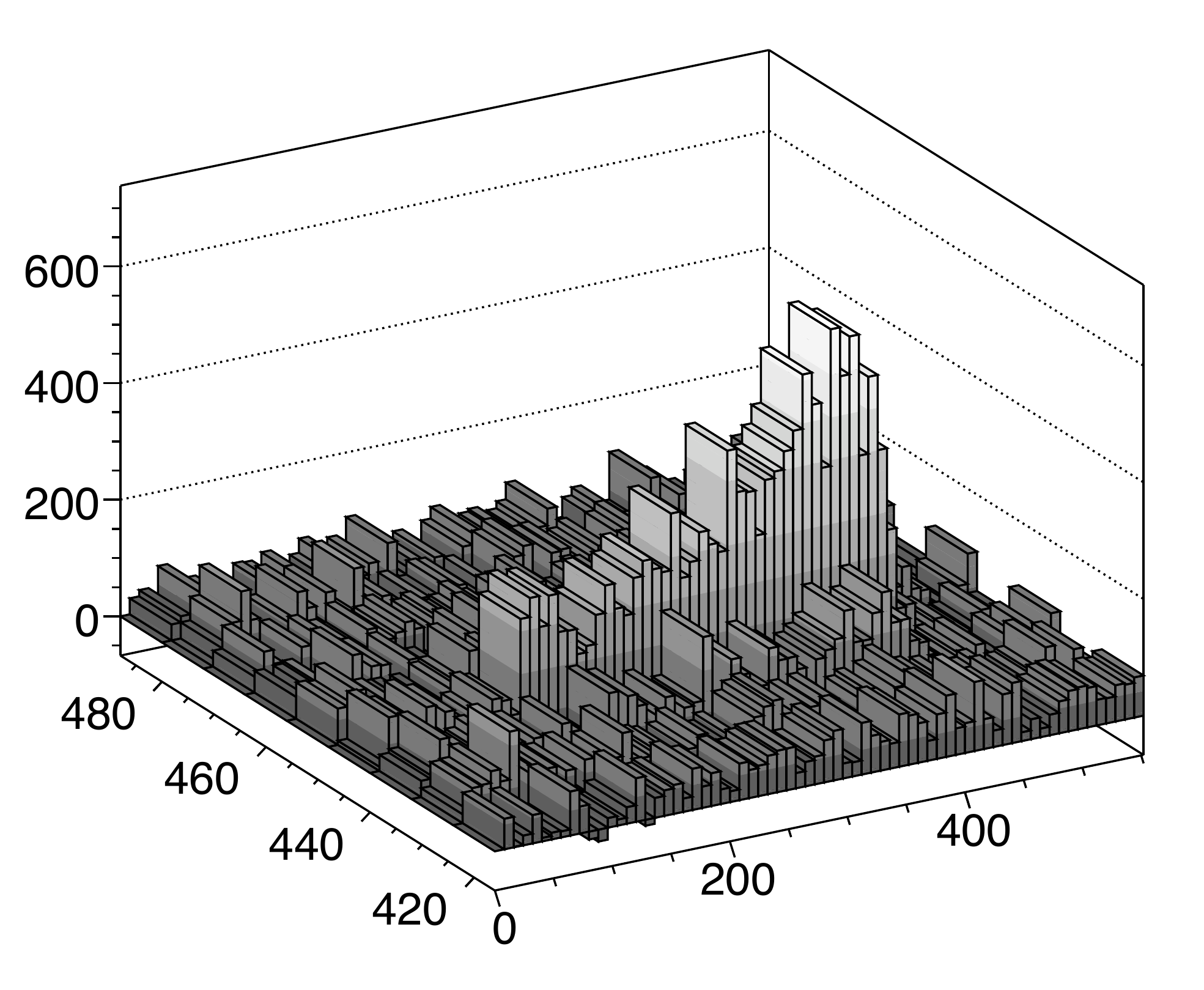} \\
\includegraphics[width=6.5cm]{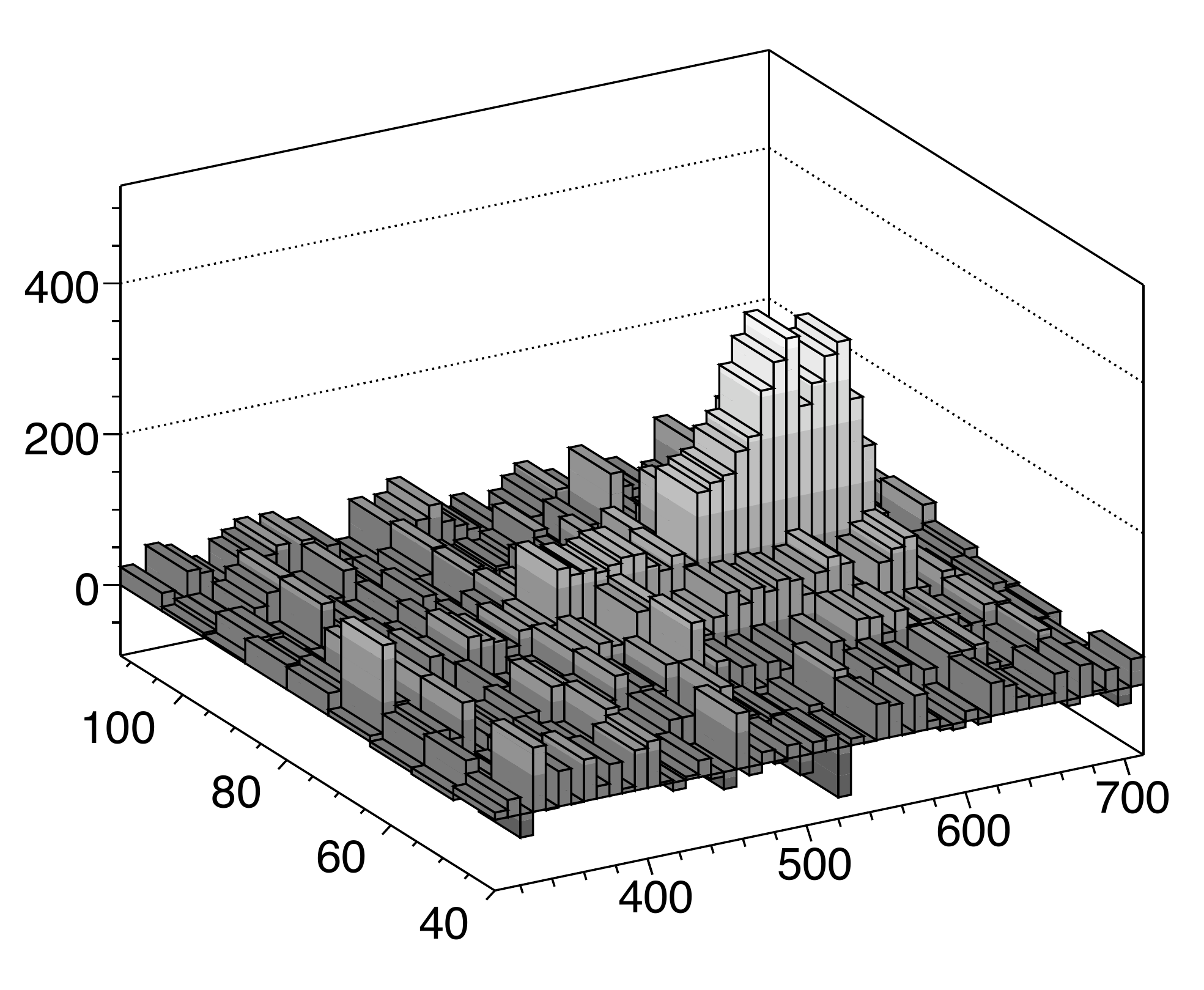} & \includegraphics[width=6.5cm]{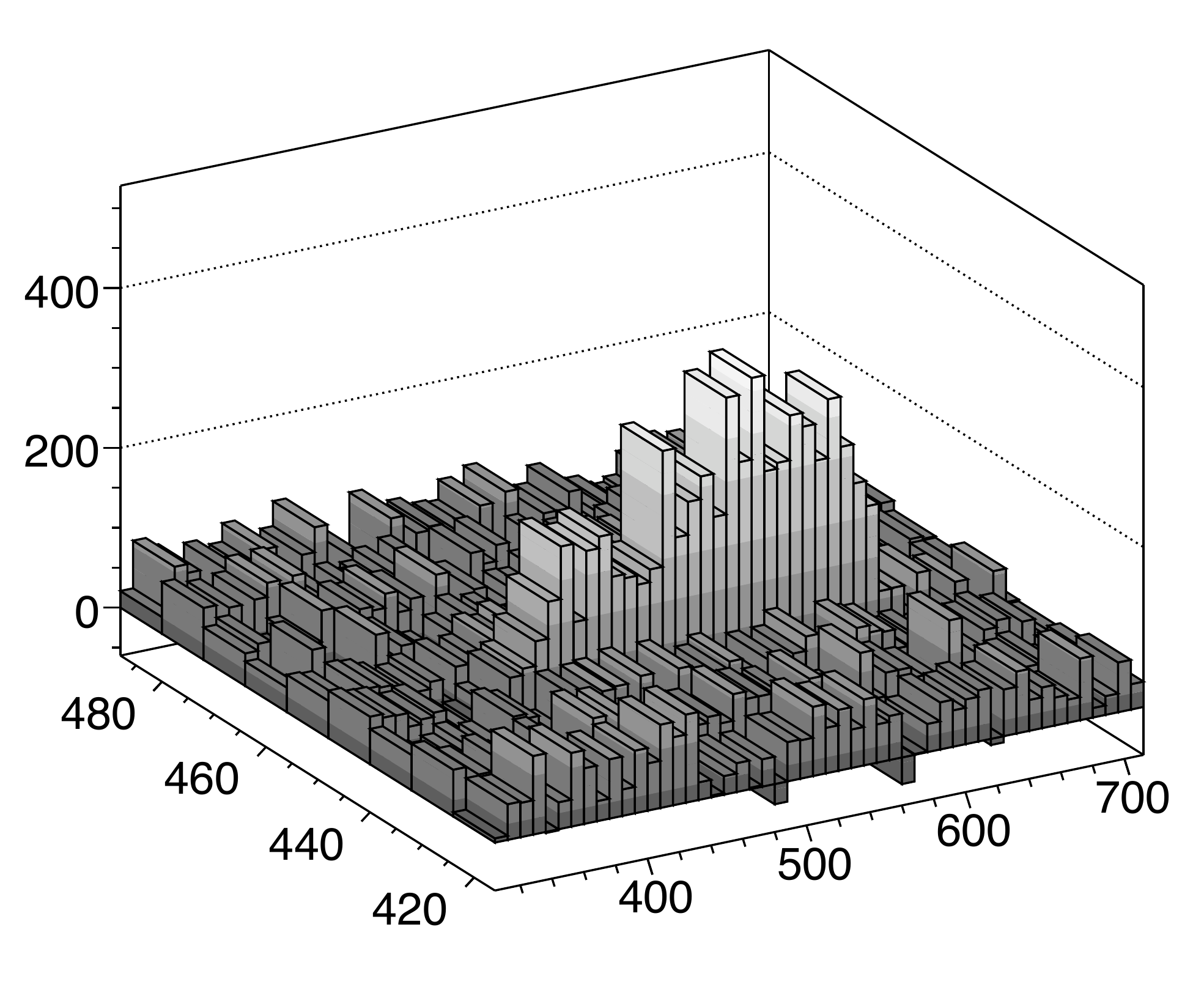} \\
\includegraphics[width=6.5cm]{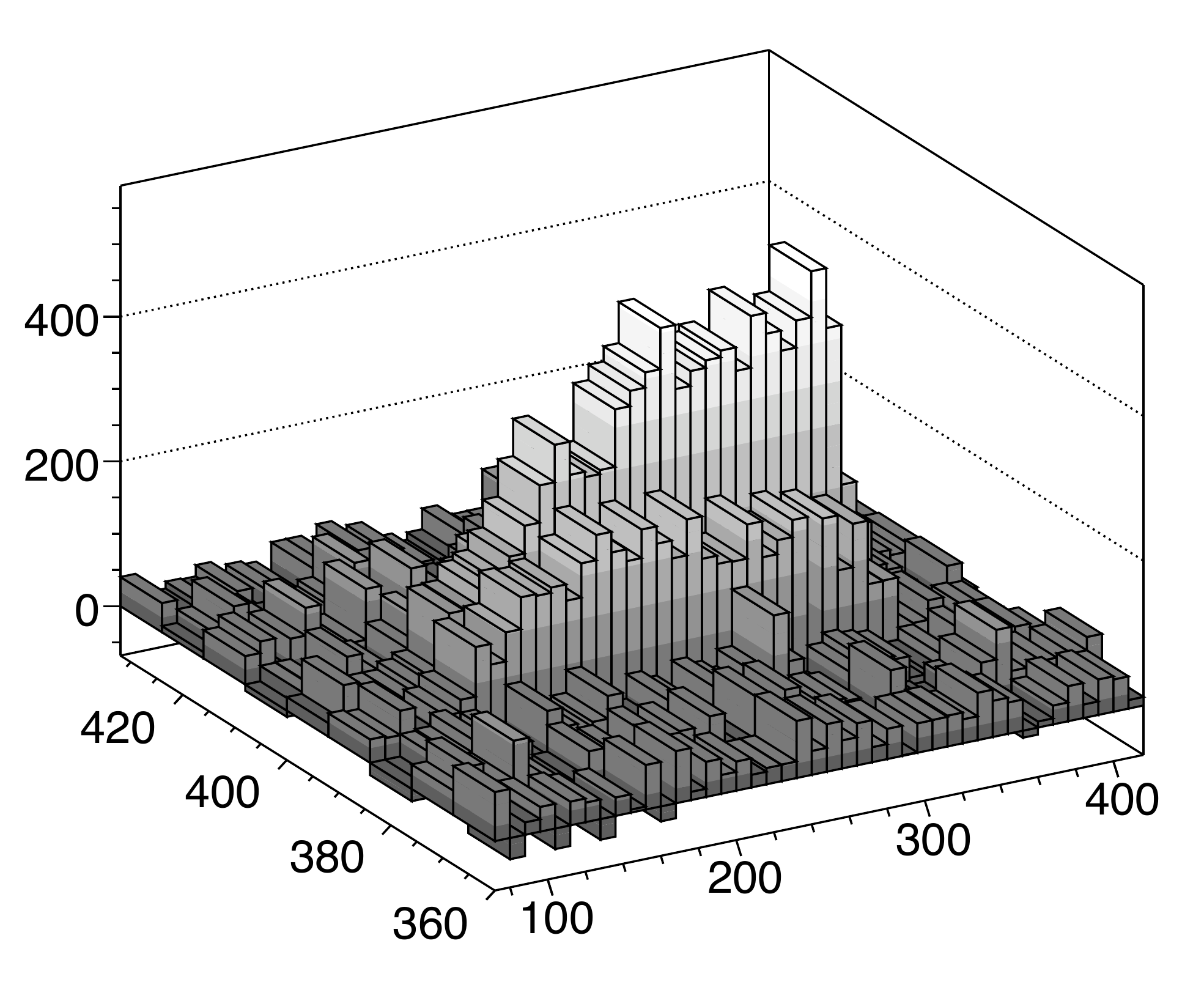} & \includegraphics[width=6.5cm]{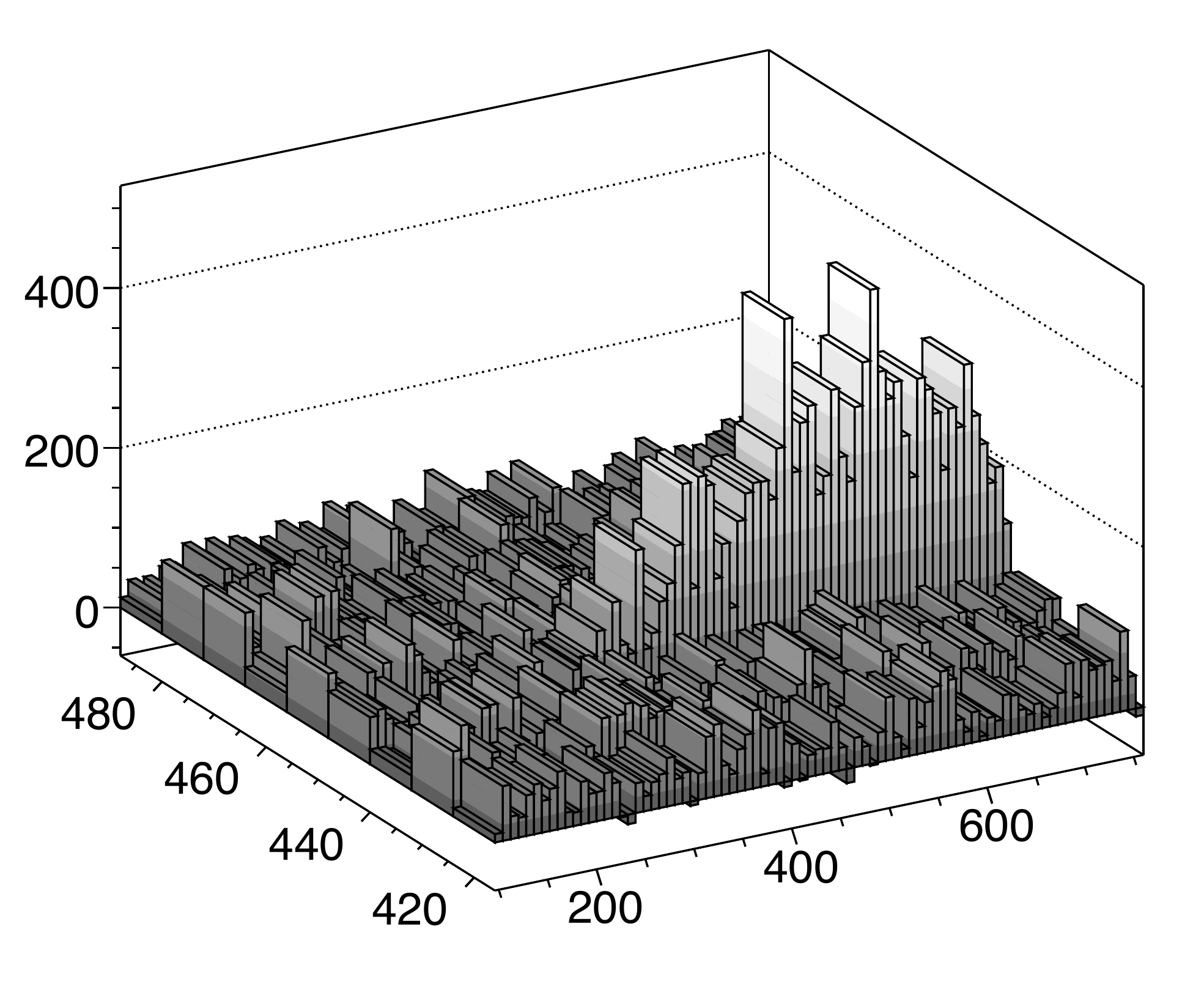}
\end{tabular}
\caption{Images of recoil tracks showing CCD coordinates and pixel
intensity. In all images  neutrons are coming from the right.
Images in the right column have the wire planes rotated by 180$^\circ$.
The noticeable asymmetry of the light yield along the wire 
indicates observation of the ``head-tail'' effect.  
\label{fg::recoil_images}}
\end{figure}

The light yield and the length of the recoil tracks are measured for all candidates that 
pass the selection criteria. 
The length is extracted by projecting the pixels in the vicinity of the wire onto the axis parallel to 
the wires and measuring the span of the pixels above the background threshold.
The light yield of the recoil segment is measured by making a projection of the CCD array to the axis perpendicular to
the direction of wires. In this projection, the recoil appears as a Gaussian signal on top of a flat background. 
The width of the Gaussian function  roughly equals the diameter of the wire, as the majority of the scintillation photons are
created around the wire where the electric field is strongest, and the mean corresponds 
to the position of a wire.
The integral of the Gaussian measures the light yield. 
Empty images that make up about 70\% of the data sample are rejected, as well as 
images that have segments shorter than 0.36~mm, images with more than one segment
per wire, and recoil tracks that fall close to the boundary of the CCD field of view. 
If a recoil is found to have scintillation light at two wires, 
only the wire with the larger scintillation signal is used in the analysis.
Approximately 5-7\% of all events pass the selection criteria. 
Some of the selected images are shown in Figure~\ref{fg::recoil_images}.
The noticeable asymmetry of the light yield along the wire 
indicates observation of the ``head-tail'' effect.

\begin{figure}[hb]
\center
\includegraphics[width=9cm]{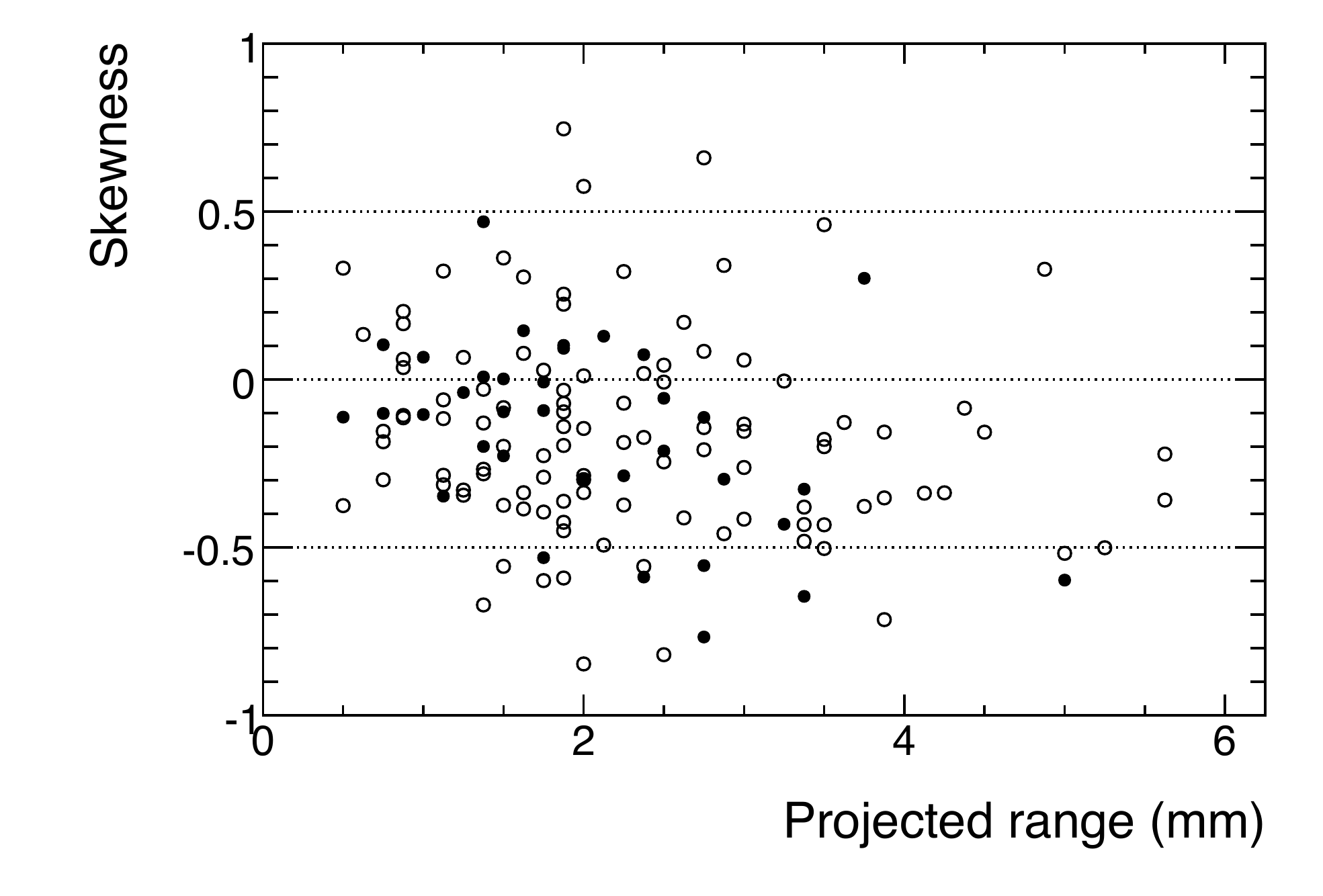}\\
\includegraphics[width=9cm]{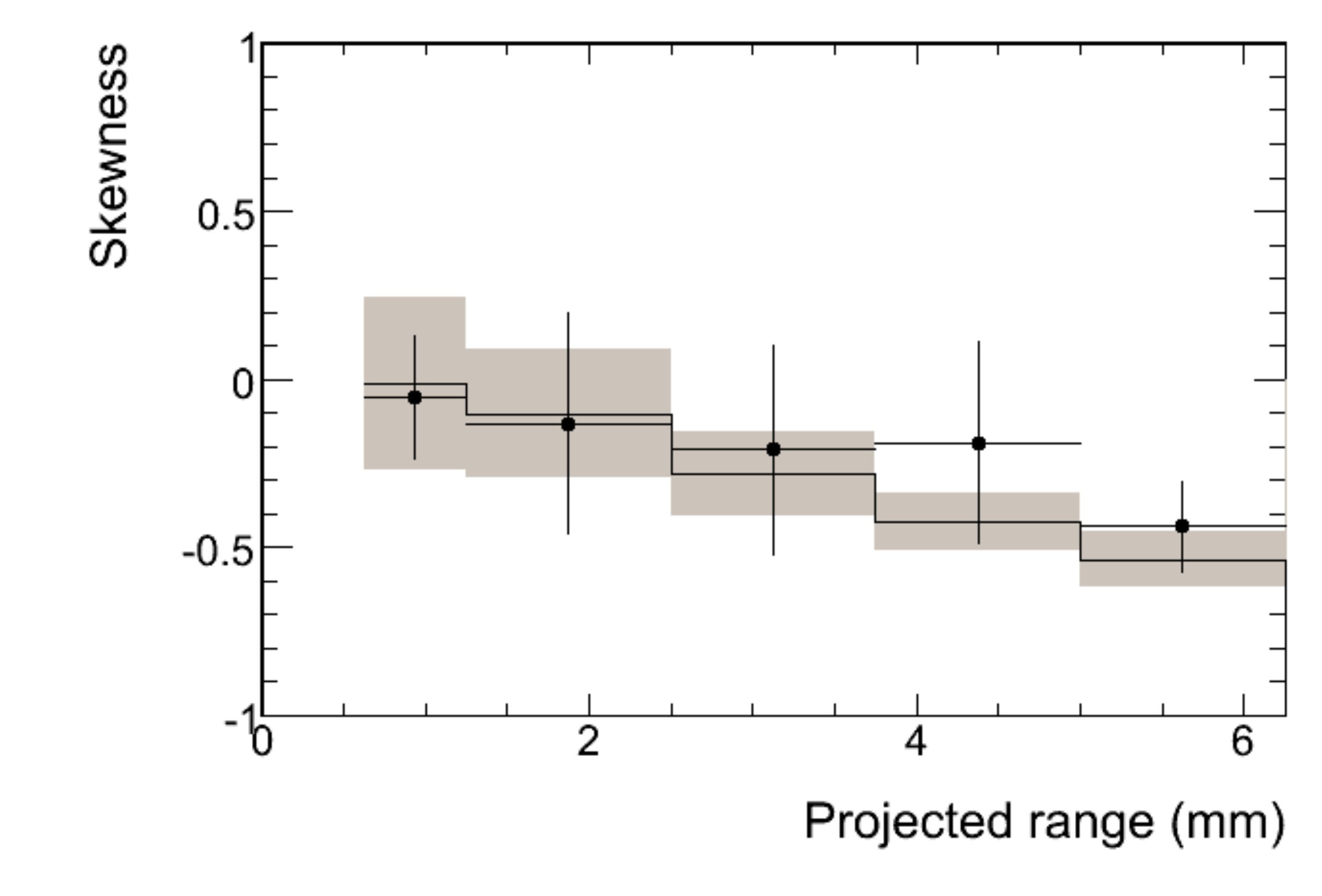} \\
\includegraphics[width=9cm]{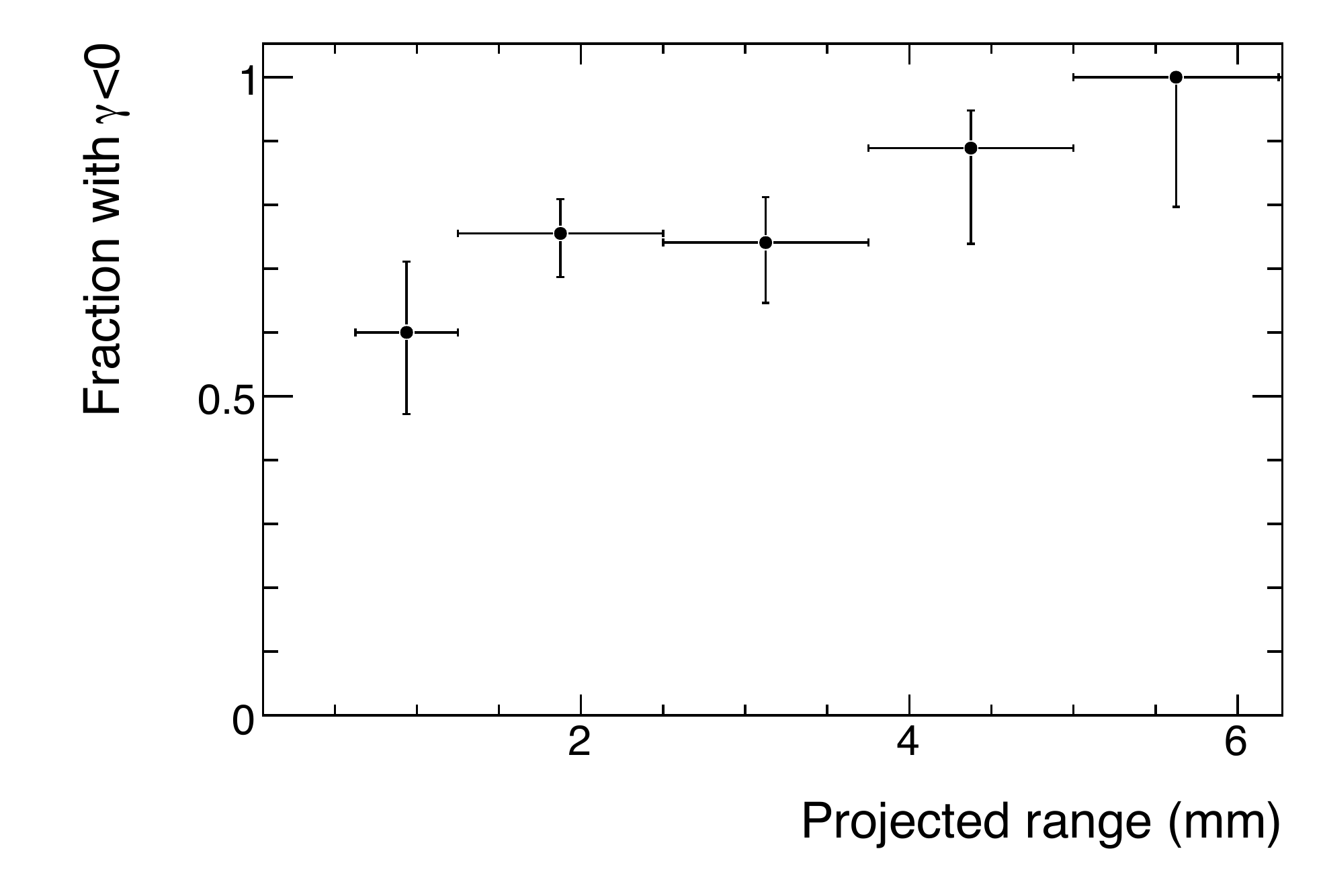}
\caption{Top plot: distribution of the skewness as a function of the track length of the recoil segments.
Open and closed circles refer to wire planes at 0$^\circ$ and 180$^\circ$ w.r.t. direction of neutrons.
Middle plot: comparison between data and simulation for the same distribution. 
The position of the dots (solid line) represents the 
mean value in each bin for data (MC). The error bars (shadowed region) measure 
the dispersion for data (MC) in each bin. 
Bottom plot: Fraction of events with negative skewness as a function of the track length.
\label{fg::recoil_energy_vs_skewness}}
\end{figure}

%
%

In order to quantify the observed asymmetry in the scintillation light,  
we define the skewness $\gamma$  as the dimensionless ratio 
between the third and second moments of the light yield along the wire coordinate ($x$): 
\begin{equation}
        \gamma = 
        \frac{\mu_3}{\mu_2^{3/2}} = \frac{ \left< (x-\left<x\right>)^3 \right> }{ \left< (x-\left<x\right>)^2 \right>^{3/2}}. 
\label{eq::skewness}
\end{equation}
The skewness $\gamma$ provides a simple measure of the ``head-tail'' asymmetry. This quantity 
is zero for perfectly symmetric distributions, and non-zero for asymmetric distributions around the mean.
The sign of the skewness indicates the slope of the light intensity along the track. 
Recoils that travel in the direction of the incoming neutrons have a decreasing light profile, and therefore 
a negative skewness.
Since the skewness is dimensionless, it is not affected 
if the coordinate is multiplied by a constant factor, which in our case is the
cosine of the recoil angle with respect to the wire.  
However, shorter segments can be  affected by the finite detector resolution.

The measured skewness as a function of the segment length is shown 
in the upper plot of Figure~\ref{fg::recoil_energy_vs_skewness}.
In ($74\pm 4$)\% of all events the skewness is measured to be negative, 
as expected for the nuclear recoils from neutrons.
 This represents a 6 $\sigma$ observation of the ``head-tail'' effect.  

Because the skewness is computed along the direction of the wires,  
the asymmetry is easier to observe for longer tracks that are better 
aligned with the anode wires and create more scintillation light. 
The bottom plot in Figure~\ref{fg::recoil_energy_vs_skewness}  shows 
the fraction of events with negative skewness as a function of the track length.

The skewness of nuclear recoils is computed using the stopping power 
and straggling according to the SRIM simulation. 
In addition, detector  effects such as diffusion and light detection inefficiencies 
are included based on the measurements obtained with $\alpha$ tracks. 
The simulation also takes 
into account that some of the recoils are only partially contained in the CCD view field.
The middle plot of Figure~\ref{fg::recoil_energy_vs_skewness} 
shows the mean values and the dispersion of the skewness in data (dots) 
and  simulation (histogram). The agreement between data and simulation is satisfactory.

\section{Discussion of Results }
\label{sec::summary}

\begin{figure}[hb]
\center
\includegraphics[width=10cm]{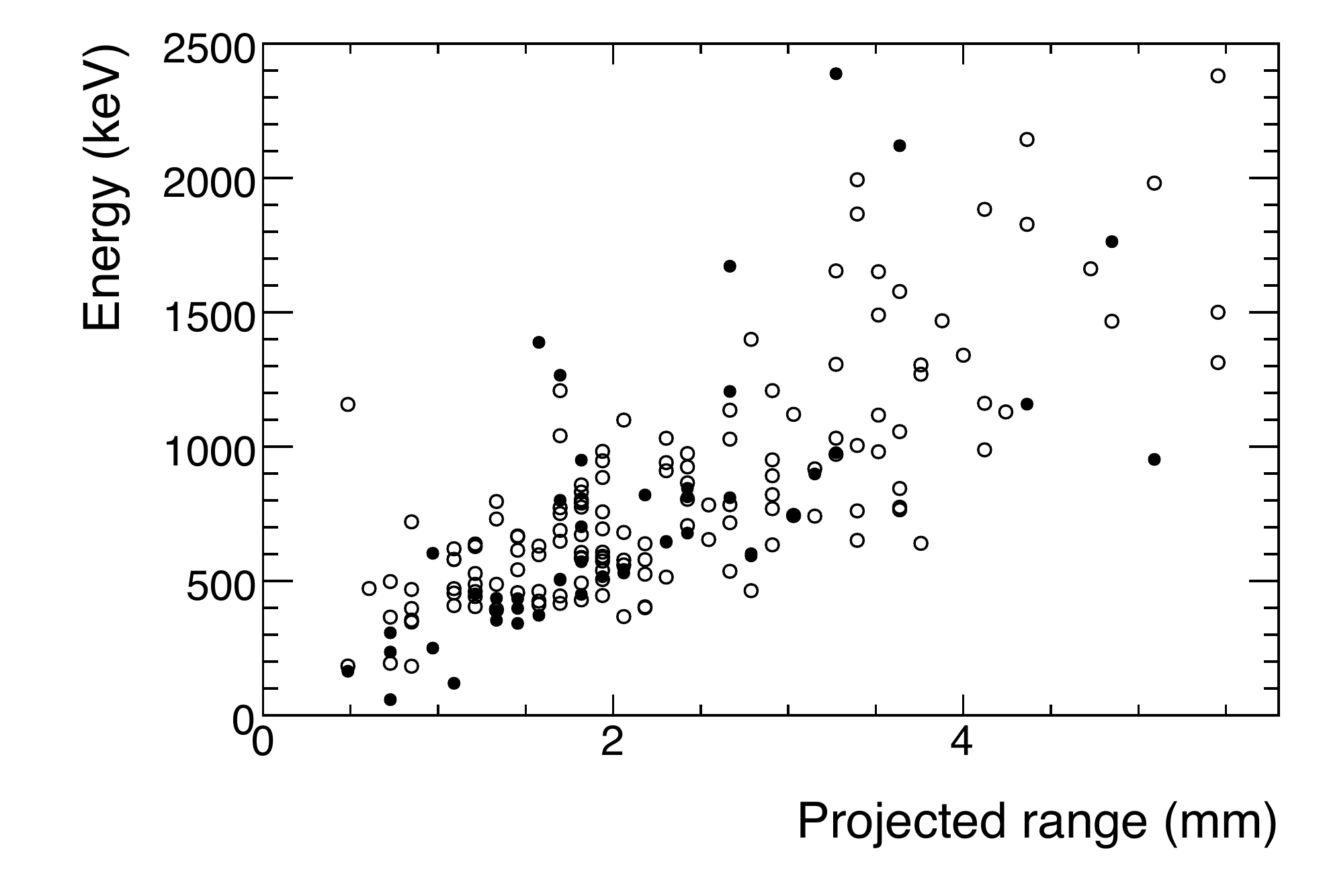}
\caption{Correlation between the energy and the range projected to the wire axis of the recoil segments in data.
Open (closed) circles refer to wire planes at 0$^\circ$ (180$^\circ$) with respect to 
the direction of the incoming neutrons.
\label{fg::recoil_energy_vs_length}}
\end{figure}

%
%
Several cross-checks are performed to validate this result.
Since the measured light yield is proportional to the energy of the recoil segment and the length is proportional to
the track range projected to the wire, these two quantities should be correlated.
Figure~\ref{fg::recoil_energy_vs_length} shows the scatter plot of the 
energy  versus projected range  of the recoil segments measured in data.  
A clear correlation is observed.

The possibility that 
the ``head-tail'' asymmetry is created by imperfections in the construction of 
the tracking chamber is excluded by taking a fraction of the data with anode wires rotated by 180 degrees with respect 
to the neutron beam. The measurement of the skewness is consistent in the two data-sets. 

We collect data with the neutron beam perpendicular to the anode wires.
In this configuration there are an equal number of recoils traveling in both directions, 
resulting in an equal number of events with positive and negative skewness.
The average skewness in all events is $\left< \gamma \right> =-0.022 \pm 0.018$,
consistent with the expected symmetric distribution. 
The average skewness due to $\alpha$ tracks traveling perpendicularly to the anodes 
is measured to be 
$\left< \gamma \right> =0.032 \pm 0.024$,  consistent with having symmetrical
scintillation transverse to the direction of the trajectory. 
The hypothesis that the scintillation signal is affected by recoil tracks leaving the drift region was discarded because 
such segments would have abrupt endings, which is inconsistent with the slowly dimming signals observed in data.

As a final check, we collect data without sources and search for signatures that resemble nuclear recoils.
In this analysis we count events with pixel yields at least five standard deviations above the background level.
We measure the rejection rate to be approximately $10^{-3}$, which can be further improved by 
taking into account the
energy, range and direction of recoil candidates.

A conservative error of 20\% was assigned to the density of the \cf4 gas to account for 
problems with the pressure measurement. 
This measurement was obtained by using  a gas-dependent thermo-couple gauge in the nuclear recoil 
measurements, which
was cross-calibrated with a more precise capacitance gauge used in the calibration measurements
with alpha particles. However, the uncertainty of the pressure measurement relates to the exponential
error on the gain measurement. 
This effect is minimized by assuming that all events are fluorine recoils, 
and by determining 
the chamber gain by adjusting the slope in Figure~\ref{fg::recoil_energy_vs_length}
to correspond to the slope of fluorine recoils. 
Using this procedure, the a gain of 8~ADC counts per keV is measured. This gain was used as 
input for MC studies and to calibrate the axes of figure~\ref{fg::recoil_energy_vs_length}. 
Other systematic errors on the energy measurement come from the non-uniformity in the gain between wires and
the stability of the gain with time.
In addition, the statistical uncertainty on the energy measurements is approximately 10\%, as 
determined from calibration with alpha particles.

The error on the recoil range comes from the non-uniformity in the wire pitch (10\%) 
and the analysis technique that overestimates
the range for low-energy recoils with the range close to the diffusion width.

\section{Conclusion and Outlook}
\label{sec::conclusion}
This study demonstrates a method for tagging the direction of low-momentum nuclear 
recoils generated by the elastic scattering of low-energy neutrons in \cf4 gas. 
The tag of the incoming particle is determined from the profile of the  
scintillation light along the track trajectory, 
leading to a 6~$\sigma$ observation of the ``head-tail'' effect for recoil energies above 500 keV.  

In the near future, these studies will be extended  to lower energy recoils produced by a 
Californium-252 source. By using a similar detector but built with higher quality standards, 
 the ``head-tail'' effect is expected to be visible down  to approximately 100~keV.  

These studies have profound implications for the development of directional dark matter detectors,
as they proves that ``head-tail'' discrimination is indeed feasible. 
Directional detectors will be essential to provide convincing evidence 
for dark matter particles in the presence of backgrounds.  

\section{Acknowledgments} 
We acknowledge support by the Advanced Detector Research Program of 
the U.S. Department of Energy (contract number 6916448), as well as the Reed Award Program, the Ferry Fund, 
the Pappalardo Fellowship program, and the Physics Department at the Massachusetts Institute of Technology.


\end{document}